# MFCCGAN: A Novel MFCC-Based Speech Synthesizer Using Adversarial Learning

*Mohammad Reza Hasanabadi, Majid Behdad, Davood Gharavian*
Shahid Beheshti University, Tehran, Iran, {m_hasanabadi, m_behdad, d_gharavian}@sbu.ac.ir

**ABSTRACT**

In this paper, we introduce MFCCGAN as a novel speech synthesizer based on adversarial learning that adopts MFCCs as input and generates raw speech waveforms. Benefiting the GAN model capabilities, it produces speech with higher intelligibility than a rule-based MFCC-based speech synthesizer WORLD. We evaluated the model based on a popular intrusive objective speech intelligibility measure (STOI) and quality (NISQA score). Experimental results show that our proposed system outperforms Librosa MFCC-inversion (by an increase of about 26% up to 53% in STOI and 16% up to 78% in NISQA score) and a rise of about 10% in intelligibility and about 4% in naturalness in comparison with conventional rule-based vocoder WORLD that used in the CycleGAN-VC family. However, WORLD needs additional data like F0. Finally, using perceptual loss in discriminators based on STOI could improve the quality more. WebMUSHRA-based subjective tests also show the quality of the proposed approach.

## 1. INTRODUCTION

Signal reconstruction is an important part of almost any system containing synthesizers [1,2] such as coding [3], text-to-speech [4], speech enhancement [5], and voice conversion systems [6]. In some cases, having features extracted and processed, recovering the original signal for posterior use is challenging. Since some feature extraction methods use nonlinear transforms, getting the signal back is not as easy as feature extraction. The limitation of bandwidth constrains the functionality of such systems. Quality and bitrate (available bandwidth) are relevant attributes that usually happen together. Often, the more the bit rate you dedicate, the more quality you get. However, available bandwidth is usually a big challenge to deal with. In such conditions, finding codecs capable to improve quality and decrease bandwidth simultaneously is of great importance. So far, communication systems incorporate several kinds of audio codecs depending on complexity, latency, bandwidth needed, and quality [7]. Most domestic PSTN centers operate at an 8 kHz sampling rate and 8-bit non-linear quantization according to ITU-T G.711, which results in 64 kbps encoding. GSM family including Full Rate (GSM-FR), Half Rate (HR), and Enhanced Full Rate (EFR), were among the first digital speech coding standards introduced to use in GSM (Global System for Mobile Communication). They operate at an average bit rate of 13 kbps, 5.6 kbps, and 12.2 kbps respectively [7]. Albeit GSM-EFR consumes less bandwidth in comparison with FR, it provides better speech quality and robustness in facing network impairments. MELPe (Mixed-Excitation Linear Predictive) vocoder algorithm, which operates at both 1.2 and 2.4 kbps is a very low-bit rate codec selected by the United States Department of Defense [8, 9]. Speex [10] is another audio compression suitable to handle VoIP and internet audio streaming. Based on the CELP algorithm [11], it operates at bit rates ranging from 2.2 to 44 kbps. Changing bit rate dynamically makes Speex among the few codecs providing Variable Bit-Rate (VBR). While traditional rule-based methods often fail to recover signals out of their features due to nonlinearities in the process of extraction, learning-based approaches present a framework to reconstruct signals better out of their features. Big data availability facilitates learning complex models. Deep neural networks play an important role in modeling complex and nonlinear behaviors. Convolutional Neural Networks (CNN), Variational Auto Encoders (VAE), and Generative Adversarial Networks (GAN) are important classes of deep learning models. These classes apply to various fields of speech processing such as speech recognition, synthesis, coding, voice conversion, and so on. Van den Oord et al. introduced WaveNet [13], a deep neural network for generating raw audio waveform. It is an autoregressive generative model, which predicts each output audio sample conditioned on some previous ones. WaveNet is based on PixelCNN, which was first introduced for image generation [14].

Although WaveNet is usually applied for text-to-speech purposes, it can adapt to coding applications. W. Bastiaan Kleijn et al. [15] suggested WaveNet generates high-quality speech from the bit stream of 2.4 kbps. C. Garbacea et al. also combined WaveNet and VQVAE [16] to produce low-bit-rate speech coding. WaveGlow is also another decoder to generate high-quality speech from Mel-spectrograms without the need for autoregression [17]. LPCNet [18], a WaveRNN [19] variant, combines linear prediction with recurrent neural networks to improve the efficiency of speech synthesis. Autoregressive models, such as WaveNet, model local structure but have slow iterative sampling and lack global latent structure. In contrast, Generative Adversarial Networks (GANs) have global latent conditioning and efficient parallel sampling [20]. MelGAN is a non-autoregressive feedforward convolutional architecture to perform audio waveform generation in a GAN setup [2]. It was almost the first work on reconstructing raw audio out of Mel-spectrograms. Due to being non-autoregressive, it is fast and suitable to be replaced as an alternative to instead of autoregressive models [2]. This paper presents a GAN-based architecture to generate raw waveform based on MFCC features (MFCCGAN). MFCC could be considered a good feature capacity to utilize in coding systems. Besides coding, CycleGAN-VC1/VC2 [21, 22]. use MFCC as conversion features for voice conversion based on WORLD vocoder [23] Therefore, better methods for MFCC inversion with higher quality and intelligibility could be also useful for improving the quality of this family of voice conversion systems. In the following sections, we take a brief review of MFCC features in section 2, then introduce our proposed GAN-based network in section 3. We investigate the perceptual optimization in section 4 and experimental results in section 5, coding applications in section 6, and finally suggest some future research directions besides the conclusion in section 7.[1]

## 2. MEL-FREQUENCY CEPSTRAL COEFFICIENTS

Mel-Frequency Cepstral Coefficients (MFCC) [24, 25] are among the state-of-the-art features in many speech processing systems. MFCCs are commonly used in speech recognition systems, which aim to detect the linguistic content of the speaker. MFCC is also used in music information retrieval applications such as genre classification [27], audio similarity measurements, etc. Discrete Cosine Transform (DCT) is the key element in MFCC to transform input utterance into output acoustic sequence. Since the MFCC algorithm applies nonlinear operators such as DCT, Logarithmic, and Mel scaling to extract coefficients, it discards lots of information by a low-rank linear projection of the Mel-spectrum and therefore, reconstructing the signal back from its MFCC features is challenging. On the other hand, MFCCs are engineered for other tasks of speech such as automatic speech recognition (ASR) [28] and speaker verification (ASV) [29]. Some attempts have been done so far for MFCC-inversion e.g.[30], Results of these researches are noteworthy, but they have followed complicated structures and used many modules which increase complexity and impose high computation costs, e.g.,[30]. As an example, an MFCC representation with n_mel=128 and n_mfcc=40 is analogous to a jpeg image with a quality set to 30% [31].

Simple reconstructions often lead to hissing sounds. Therefore, we propose to learn the conversion process through the deep neural network with a simple model, without any extra module to generate a more natural time-domain raw waveform. We have used an adversarial-based setup introduced in [2] and conditioned on MFCC features as input and called it MFCCGAN. Since adversarial models represent an implicit distribution density, such models are less exposed to over-smoothing which is a common challenge in speech tasks such as coding, synthesis, and conversation. In the following, we investigate each part of the architecture.

## 3. NETWORK ARCHITECTURE

### 3.1. Generator

The generator of the MFCCGAN speech synthesizer is a fully convolutional network. A stack of transposed convolutional layers has been used to upsample the input MFCC from 256x lower temporal resolution to a raw waveform. Following each convolutional layer, a block of residual layers with a dilation attribute is put on. Dilation leads to a greater receptive field of input, which could better model the variations. The upsampling applies in four stages: two 8x and two 2x resolution increases. 1D transposed convolution operator is applied in each stage to increase the resolution. Residual stacks do not change the input size. Each residual stack consists of three convolutional layers with different dilation parameters. As it is mentioned earlier, these stacks look at the input with a greater field with the aid of dilation properties. Table 1 shows the specifications of each layer of the generator in sequence.

### 3.2. Discriminator

Adopting a multi-scale architecture containing three identical discriminators, which operate on different audio scales (sampling rates), is an important feature of discriminators. Downsampling a signal into lower resolution lets us look into the signal with different frequency resolutions. Higher-frequency behavior cues more at higher scales while a lower sampling rate could reflect the low-frequency behavior. Therefore, scaling the output signal into three levels could lead to better judging of the quality.

The downsampling process inside each discriminator is just done using convolutional layers. No pooling is adopted in discriminators except in the input sequence. The structure of all three discriminators is identical, so, only the attribute details of discriminator I, are provided in table 2.

The attributes of each convolution layer of discriminators like kernel, stride, padding, and dilation are adjusted so that the downsampled output signal of each discriminator is according to table 3.

### 3.3. Training objective

Inspiring by [2], we have used the Least Squares GAN [32] based on equations (1) to (4) as objective functions::

$$L_D = \min_{D_k} \left( E_x \left[ 1 - D_k(x) \right]^2 + E_s \left[ 0 - D_k(G(s)) \right]^2 \right) \quad (1)$$
$$\forall k = 1, 2, 3$$

$$L_G = \min_G E_s \left[ \sum_{k=1,2,3} \left[ 1 - D_k(G(s)) \right]^2 \right] \quad (2)$$

$$L_{FM}(G, D_k) = E_{x,s \sim p_{data}} \left[ \sum_{i=1}^{T} \frac{1}{N_i} \left\| D_k^{(i)}(x) - D_k^{(i)}(G(s)) \right\|_1 \right] \quad (3)$$

$$L_G = \min_G \left( E_s \left[ \sum_{k=1,2,3} \left[ 1 - D_k(G(s)) \right]^2 \right] + \lambda \sum_{k=1,2,3} L_{FM}(G, D_k) \right) \quad (4)$$
$$\forall k = 1, 2, 3$$

Equation (1) pushes three discriminators to discriminate strictly between real and fake reconstructed data. On the other hand, (2) drives a generator to generate raw waveform similar to the real data, to cheat the discriminator. Since GAN networks are prone to collapse modes, we also utilize an extra loss function LFM as (3) named feature match [2] to trace the status of training. Each discriminator outputs feature maps at its consecutive layers with different time resolutions; therefore, these extra features are used to form the feature match loss between original and predicted samples. The final objective of the generator is then termed (4).

## 4. PERCEPTUAL OPTIMIZATION USING STOI

To increase the quality of synthesized speech more, inspired by MetricGAN [33], we tried to force the discriminators to learn to judge between real and fake data based on a perceptual intelligibility metric, STOI [34]. Therefore, we calculated STOI between real and fake training utterances in each batch and substituted it instead of zero in (1). So, we finally used (5) as a novel perceptual loss function instead of (1) in discriminators to increase the intelligibility of generated utterances of the generator whereas experimental results verified this idea :

$$L_D = \min_{D_k}\Big(E_x\big[1-D_k(x)\big]^2 + E_s\big[STOI(x,G(s)) - D_k(G(s))\big]^2\Big)$$
$$\forall k = 1,2,3$$
(5)

---

[1] *Source code is available at* *https://github.com/MohammdReza2020/mfccgan*

Table 1. Specifications of each layer of Generator in sequence

| No. | Layer | Input-size | In-channel | Out-channel | Kernel | Stride | Padding | Dilation | Output-size |
|---|---|---|---|---|---|---|---|---|---|
| 1 | Conv1d | N×39×32* | 80 | 512 | 7 | 1 | 0 | 1 | N×512×32 |
| 2 | Conv1d_transpose(8x) | N×512×32 | 512 | 256 | 16 | 8 | 4 | 1 | N×256×256 |
| 3 | Residual stack | N×256×256 | 256 | 256 | 1-3** | 1 | 0 | 0-3-9*** | N×256×256 |
| 4 | Conv1d_transpose(8x) | N×256×256 | 256 | 128 | 16 | 8 | 4 | 1 | N×128×2048 |
| 5 | Residual stack | N×128×2048 | 128 | 128 | 1-3 | 1 | 0 | 0-3-9 | N×128×2048 |
| 6 | Conv1d_transpose(2x) | N×128×2048 | 128 | 64 | 4 | 2 | 1 | 1 | N×64×4096 |
| 7 | Residual stack | N×64×4096 | 64 | 64 | 1-3 | 1 | 0 | 0-3-9 | N×64×4096 |
| 8 | Conv1d_transpose(2x) | N×64×4096 | 64 | 32 | 4 | 2 | 1 | 1 | N×32×8192 |
| 9 | Residual stack | N×32×8192 | 32 | 32 | 1-3 | 1 | 0 | 0-3-9 | N×32×8192 |
| 10 | Conv1d | N×32×8192 | 32 | 1 | 7 | 1 | 0 | 1 | N×1×8192 |

\* N indicates the batch size.
\** In residual blocks, three convolutional layers with kernel sizes of one, two, and three are applied.
\*** In residual blocks, three convolutional layers with dilation sizes of $3^j$ (j=0, 1, 2) are applied.

Table 2. Specifications of each layer of Discriminator in sequence

| No. | Layer | Input-size | In-channel | Out-channel | Kernel | Stride | Padding | Dilation | Output-size |
|---|---|---|---|---|---|---|---|---|---|
| 0 | input | N×1×8192 | 1 | 1 | - | - | - | - | N×1×8206 |
| 1 | Conv1d | N×1×8206 | 1 | 16 | 15 | 1 | 0 | 1 | N×16×8192 |
| 2 | Conv1d (downsample 4x) | N×16×8192 | 16 | 64 | 41 | 4 | 20 | 1 | N×64×2048 |
| 3 | Conv1d (downsample 4x) | N×64×2048 | 64 | 256 | 41 | 4 | 20 | 1 | N×256×512 |
| 4 | Conv1d (downsample 4x) | N×256×512 | 256 | 1024 | 41 | 4 | 4 | 1 | N×1024×128 |
| 5 | Conv1d (downsample 4x) | N×1024×128 | 1024 | 1024 | 41 | 4 | 20 | 1 | N×1024×32 |
| 6 | Conv1d | N×1024×32 | 1024 | 1024 | 5 | 1 | 2 | 1 | N×1024×32 |
| 7 | Conv1d | N×1024×32 | 1024 | 1 | 3 | 1 | 1 | 1 | N×1×32 |

Table 3. Input and output size of each Discriminator

| Discriminator No. | Input-size | Output-size |
|---|---|---|
| Discriminator I | N×1×8192 | N×1×32 |
| Discriminator II | N×1×4096 | N×1×16 |
| Discriminator III | N×1×2048 | N×1×8 |

## 5. EXPERIMENTAL RESULTS

To evaluate the proposed idea, we carried out two categories of assessments: objective and subjective tests. For the objective test, we selected the short-time objective intelligibility (STOI) measure which is a popular metric for the assessment of speech intelligibility [33]. There are different aspects of speech quality and among them, intelligibility and naturalness are the most important quality aspects. STOI metric has a higher correlation to subjective listening tests than simple L1 or L2, SNR, segmental SNR, and many other measures [33]. Therefore, this metric can be a good choice for the performance assessment of our proposed system, because the reconstructed and reference speech signals are time-aligned. We have also used a new no-reference naturalness objective assessment representing subjective scores called NISQA [35]. NISQA tool is a five-scaled (0-5) neural-based and non-intrusive objective quality measure for assessing naturalness. This version of NISQA (NISQA-TTS) is trained end-to-end and the time-dependency modeling and time-pooling are achieved through a Self-Attention mechanism customized for TTS applications. Besides the referenced and no-referenced objective tests, we did a simplified crowdsourced subjective test using WebMUSHRA [36]. In this experiment, we used the LJSpeech dataset [37] to prepare the sample speech utterances for training, validation, and test set. We have used 13000 utterances of the LJSpeech dataset to train the model and 100 samples (non-overlapping) utterances to evaluate our proposed method for MFCC-inversion. We used STOI and NISQA measures for assessing and comparing the intelligibility of our proposed system outputs with results of a rule-based MFCC-inversion tool, WORLD, as a strong reference system that is a high-quality vocoder and widely regarded as a state-of-the-art model in SPSS [38]. To consider the dimensionality of MFCC features, we have investigated different MFCC numbers. Table 4 shows the final results in terms of the STOI measure which is defined in a one-third octave band domain. As it is illustrated, the STOI score that is averaged on 100 utterances reconstructed by our proposed system is significantly higher than that of the conventional and rule-based MFCC-inversion WORLD (about 10% higher). For these experiments, we used 36 coefficients of MFCCs for inversion in GAN-based architecture and also for traditional WORLD MFCC-inversion. Although WORLD needs further data such as fundamental frequency and also voiced/unvoiced decision procedure, our proposed method gains higher quality in terms of objective tests. We measured and noted the STOI of synthesized speech in table 4 with MelGAN by 80 Mel-spectrogram coefficients as a reference. According to this table, at least a 4% increase in the NISQA score could be inferred. To better specify the performance of our proposed system, we compare our proposed system with its rival, Librosa as a well-known MFCC-inversion tool. Meanwhile, we used Librosa for MFCC feature extraction to prepare input for training and evaluating our proposed model, we planned another experiment for comparison between our proposed model and Librosa as a rule-based conventional MFCC feature extractor and MFCC-inversion tool. We extracted MFCC coefficients for 100 original speech utterances selected from the LJspeech database. After that, we reconstructed 100 speech files from an MFCC sequence of 100 original files by Griffim-Lim (Librosa) as a synthesis tool. We measured the intelligibility of these 100 reconstructed files by STOI metric in comparison with their 100 paired original utterances (files). Table 5 shows these results. We repeated this experiment in five different numbers of MFCC features 13, 24, 36, 39, and 80 for each of the 100 files. As can be observed in this table, our proposed system outperforms the Librosa MFCC-inversion obviously in terms of intelligibility by 26% up to 53% increase in STOI and naturalness by 16% up to 78% increase in NISQA score.

Table 4. STOI/NISQA measures for reconstructed signals by the proposed model in comparison with WORLD and MelGAN Speech Synthesizers

| Synthesis Tools & Number of Features | | |
|---|---|---|
| **WORLD Synthesizer** MFCC#36 | **MFCCGAN** MFCC#36 | **MelGAN** Mel-Spectrogram Coefs #80 (As Reference) |
| 0.6911 | 0.7664 | 0.9413 |
| 2.9501 | 3.0777 | 3.1934 |

Table 5. Intelligibility (in terms of STOI/NISQA scores) for the reconstructed speech by our proposed model in comparison with Librosa MFCC-Inverted speech utterances

| MFCC-Inversion Method | | MFCC Feature Dimensionality | | | | |
|---|---|---|---|---|---|---|
| | | #13 | #24 | #36 | #39 | #80 |
| | Librosa MFCC-Inversion | 0.4415 1.6227 | 0.4688 2.3964 | 0.5408 2.6485 | 0.5537 2.6920 | 0.6165 3.0824 |
| | Our Proposed System | 0.6779 2.9041 | 0.7122 2.9088 | 0.7664 3.0777 | 0.7674 3.2092 | 0.7821 3.5845 |

To compare our proposed system with Librosa, and WORLD subjectively; we carried out a subjective webMUSHRA experimental test. We performed assessments, following the ITU recommendations [39]. As it is illustrated in Fig.1, the proposed approach gives good-quality results. It is worth noting that MFCCGAN quality is close and even higher than WORLD although WORLD gets extra information such as pitch and Voiced/Unvoiced frames while MFCCGAN is only dependent on the MFCC features. Such a system can be used for various areas of speech processing e.g., speech synthesis, speech coding, voice conversion, and other similar speech reproduction applications, wherever the MFCCs have been used as features. To evaluate the effects of (5), we trained again our proposed system MFCCGAN by #36 MFCCs using a novel loss function. In this new configuration, the STOI score increased to 0.7764 (1.3% increase), and the NISQA score to 3.4027 (10% increase of naturalness) in comparison with ordinary MFCCGAN.

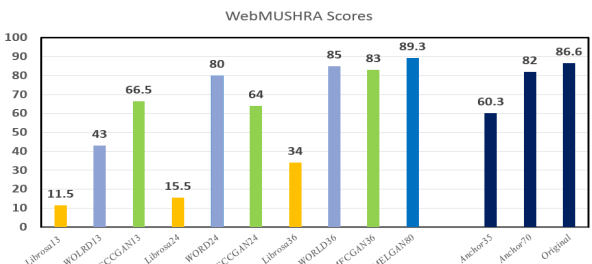

Fig.1. Results of WebMUSHRA Subjective Test

## 6. CODING APPLICATION

Since MFCCGAN extracts certain numbers of features at each frame, this system can be used for high-quality speech coding. Depending on the available bandwidth and the desired quality, according to Table 6, bit-rates ranging from 13 to 80 kbps can be achievable. Our future work will focus on reaching a high quality and low bit rate audio coding based on MFCCGAN.

Table 6. Coding Bitrates

| No. MFCC | 13 | 24 | 36 | 39 |
|---|---|---|---|---|
| Bitrate with 8-bit (kbps) | ≈13 | ≈24 | ≈36 | ≈40 |
| Bitrate with 16-bit (kbps) | ≈26 | ≈48 | ≈72 | ≈80 |

## 7. CONCLUSION AND FUTURE WORKS

In this paper, we proposed an MFCC-inversion-based speech synthesizer using adversarial learning. The results show that the proposed approach improves the intelligibility measure (STOI) by at least 10% and naturalness by 4% in comparison with a traditional rule-based algorithm WORLD. Also, the proposed system increases by at least 26% intelligibility and 16% naturalness in comparison with Griffin-Lim (Librosa tool). Subjective test results of the proposed system also outperform its conventional counterparts. This work could be used as a synthesizer in many speech tasks such as coding, text-to-speech, and voice conversion applications. Applying a perceptual (STOI) optimization could finally improve the output quality some more. Since the proposed approach achieved a higher webMUSHRA score at low bit rate coding, it hints that such an approach could be improved to be applied to low-bit rate audio coding. In addition to working on quality improvement, our future work will focus on reaching a high quality and low bit rate audio coding based on MFCCGAN. MFCCGAN can also be adopted in speech enhancement tasks. Making MFCCGAN end-to-end is also our future research prospect.